# Parameterization of the phase shift $\delta_{33}$ for $\pi N$ Scattering

Mohamed E. Kelabi[1]

**Abstract**
From the partial wave analysis, the phase shift $\delta_{33}$ of pion nucleon scattering containing the $\Delta(1232)$ resonance, corresponding to isospin $I=3/2$ and angular momentum $J=3/2$, has been parameterized over the energy range $1100 < W < 1375$ MeV, using $p\pi^+$ data. The result of our parameterization shows good agreement in comparison with the available experimental data.

## 1.1 Introduction

Scattering can change both amplitudes and phases of the outgoing waves. The change is commonly expressed in terms of the $l^{th}$ phase shift and an inelasticity parameter[1], [2]

$$\eta_l = \rho_l \, e^{2i\delta_l} \qquad (1.1)$$

where the absorption parameter $\rho_l$ is real and ranges between $0 < \rho_l < 1$. For elastic scattering $\rho_l = 1$, Eq. (1.1) becomes

$$\eta_l = e^{2i\delta_l} = 1 + 2i \sin \delta_l \, e^{i\delta_l}. \qquad (1.2)$$

If we insert Eq. (1.2) into the scattering amplitude,

$$f_l = \frac{\eta_l - 1}{2iq},$$

we obtain

$$f_l = \frac{\sin \delta_l \, e^{i\delta_l}}{q} \qquad (1.3)$$

where $q$ is the momentum of pion in the CM frame,

---

[1] Physics Department, Al-Fateh University, Tripoli, LIBYA.

$$q^2(W) = \frac{\left[W^2 - (m_N + m_\pi)^2\right]\left[W^2 - (m_N - m_\pi)^2\right]}{4W^2} \quad (1.4)$$

here $m_N$ and $m_\pi$ are the nucleon and pion rest masses, respectively, and

$$W^2 = s = m_N^2 + 2m_N E_\gamma \quad (1.5)$$

is the total CM energy for pion photoproduction; $E_\gamma$ is the photon energy in the Lab system. The Mandelstam variable $s = (P_1 + K)^2 = (P_2 + Q)^2$ is defined by the convention given in Fig. 1.

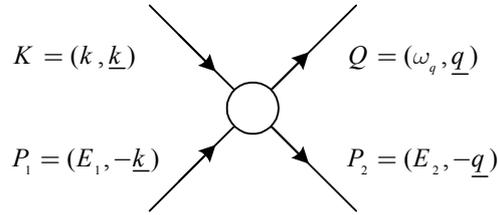

$K = (k, \underline{k})$  $Q = (\omega_q, \underline{q})$
$P_1 = (E_1, -\underline{k})$  $P_2 = (E_2, -\underline{q})$

Fig. 1. The $s$ channel scattering for $\pi N$.

## 1.2 Theory and formalism

The inverse of the scattering amplitude Eq. (1.3) reads

$$\frac{1}{f_l} = K_l^{-1} - iq \quad (1.6)$$

where we have introduced

$$K_l^{-1} = q \cot \delta_l, \quad (1.7)$$

which is an analytic function of $W^2$ and hence $q^2$ at threshold. To investigate the energy dependence of the resonant amplitude, we define

$$H(W) = q^{2l+1} \cot \delta_l = q^{2l} K_l^{-1}. \quad (1.8)$$

This is important for resonances near threshold, since that provided $H(W)$ is finite at threshold and, it will automatically give the correct threshold behaviour for $K_l^{-1}$, and so for the scattering amplitude[2]. To locate the pole position properly, we constrict on accurate parameterization of the phase shift data. Specifically we consider the $p$-wave parameterization:

$$q^3 \cot \delta = a_0 + a_2 q^2 + a_4 q^4 + a_6 q^6 + a_8 q^8 \qquad (1.9)$$

where the $\pi N$ phase shift $\delta \equiv \delta_{2I\,2J} = \delta_{33}$, as measured in $p\pi^+$ scattering.

**1.3 Results**

In Fig. 2, we plot our parameterization of the $\delta_{33}$ compared with the available $p\pi^+$ experimental data[3].

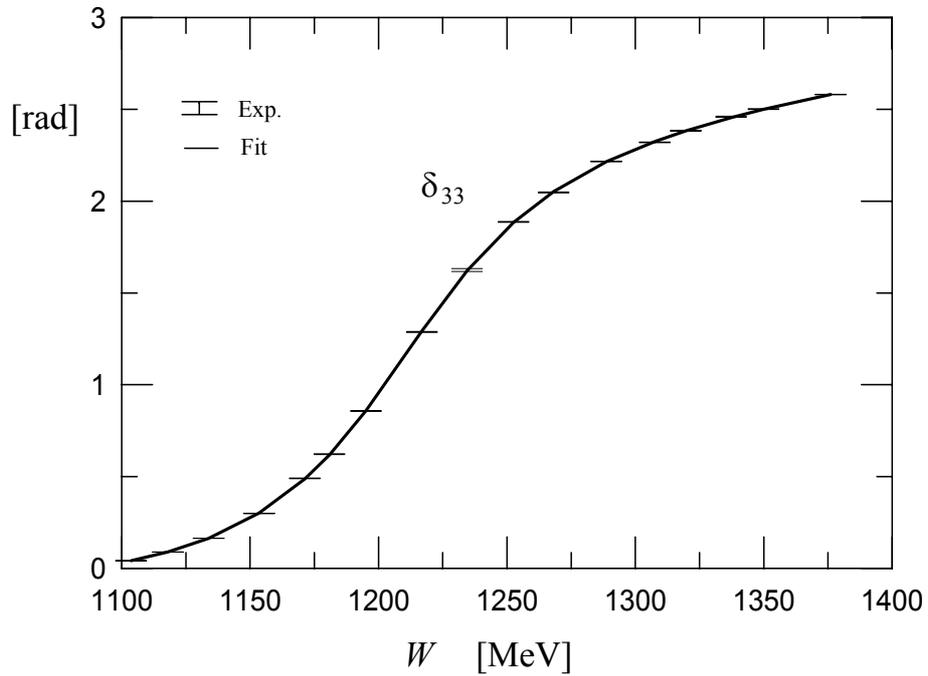

Fig. 2. Phase shift parameterization of the $\delta_{33}$. The value of error on each data point is presented by the symbol ⊥.

The fit is based on the chi-squared per degrees of freedom,

$$\chi^2 \equiv \frac{1}{m-p} \sum_m \left( \frac{exp_m - fit_m}{\varepsilon_m} \right)^2$$

where $m$, $p$, and $\varepsilon$ are the number of data points, degrees of freedom, and the corresponding error on each data point, respectively. Giving the fitting parameters

$$a_0 = 5.0504$$
$$a_2 = -0.43566$$
$$a_4 = -0.56345$$
$$a_6 = 0.01781$$
$$a_8 = -6.35254 \times 10^{-3}$$

in charged pion mass units, with chi-squared per degrees of freedom $\chi^2 = 3.4$.

### 1.4 Conclusion

The given parameterization shows a good fit compared with the experiment of $p\pi^+$, over the energy range $1100 < W < 1375$ MeV. The obtained phase shift $\delta_{33}$ can be used to locate the pole position of the scattering amplitude $f_{33}$. It can also be employed, using appropriate techniques, to extract the position and width of the $\Delta(1232)$ resonance. Further improvement of the fit can be achieved by adding more terms.

**Acknowledgement**
I am grateful to Dr. Graham Shaw[2] for his valuable comments.

---

[2] Department of physics and Astronomy, University of Manchester, Manchester, UK.